\documentclass[useAMS,usenatbib,onecolumn]{mn2e}

\usepackage{epsfig}
\usepackage{url}

\usepackage[version=3]{mhchem}

\usepackage{color}
\usepackage{natbib}

\newcommand{\leftexp}[2]{{\vphantom{#2}}^{#1}{#2}}   %\leftexp{e}{F} produces a left superscript eF

\newcommand{\cm}{cm$^{-1}$}

\newcommand{\p}{^\prime}
\newcommand{\pp}{^{\prime\prime}}
\newcommand{\ai}{\textit{ab initio}}

\title[Exomol molecular line lists VI: PN]{Exomol molecular line lists VI: A high temperature line list for Phosphorus Nitride}
\date{\today}
\author[Yorke et al]{Leo Yorke, Sergei N. Yurchenko,  Lorenzo Lodi and Jonathan Tennyson \\
Department of Physics and Astronomy, University College London, Gower Street, WC1E 6BT London, UK}

\date{Accepted XXXX. Received XXXX; in original form XXXX}

\begin{document}

\maketitle

\begin{abstract}

  Accurate rotational-vibrational line lists for $^{31}$P$^{14}$N and $^{31}$P$^{15}$N
  in their ground electronic states are computed.  The line lists are produced using an
  empirical potential energy curve obtained by fitting to the experimental transition
  frequencies available in the literature in conjunction with an accurate, high level
  \textit{ab initio} dipole moment curve. In these calculations the programs DPotFit
  and LEVEL~8.0 were used. The new line lists reproduce the experimental wavenumbers
  with a root-mean-square error of 0.004~cm$^{-1}$. The line lists cover the frequency
  range 0--51000 cm$^{-1}$, contain almost 700~000 lines each and extend up to a
   maximum vibrational level of $v$=66 and a maximum rotational level of $J$=357. They
  should be applicable for a large range of temperature up to, at least, 5000~K. These
  new line lists are used to simulate spectra for PN at a range of temperatures and are
  deposited in the Strasbourg data centre. This work is performed as part of the ExoMol project.
\end{abstract}

\textit{molecular data; opacity; astronomical data bases: miscellaneous; planets and satellites: atmospheres; stars: low-mass}

%\newpage

%\tableofcontents

%\newpage

\section{Introduction}

Phosphorus nitride, PN, is a stable, strongly bound, closed shell
diatomic molecule. Its electronic ground state has $^1\Sigma^+$
symmetry and its dissociation energy is 51940$\pm$170~\cm{}
\citep{69Gixxxx.PN,06CaClLi.PN}. It has no low-lying electronically
excited states: the lowest excited state is the  a~$\! ^3\Sigma^+$ state at
about 25000~\cm{} (for near-equilibrium geometries, $\approx$1.5~\AA)
while the lowest-lying singlet is the  A~$\! ^1\Pi$ state at about 40000~\cm{}
\citep{83GrKaxx.PN,93deFeKo.PN}.

PN is an astrophysically  important molecule used to probe different regions of the
interstellar medium (ISM). PN was first observed in the ISM by \citet{87Zixxxx.PN}, where it
was noted that PN was more abundant than originally predicted.  It was also observed in the warm
star forming regions Ori (KL), W51M, Sgr B2 by \citet{87TuBaxx.PN} as well as %, and it was further
% observed
in warm molecular clouds of the star forming regions M17SW and DR~12OH by
\citet{90TuTsBa.PN}, suggesting high temperature chemistry is important for its
formation. PN was detected in the carbon-rich source CRL 2688 by \citet{08MiHaTe.PN}.
Most recently it was observed in the Lynds 1157 B1 shocked region by \citet{11YaTaSa.PN} and in
the outflow from the protostar IRAS 20386+6751 by \citet{12YaTaWa.PN}. The chemical
mechanism of the PN formation was recently studied by \citet{09ViPeMa.PN}. As yet PN has not been
observed in hotter bodies; however
\citet{06ViLoFe.H2S} studied phosphorous and found
that in hotter objects, such as brown dwarfs,
   P-bearing gases such as PN become increasingly
   important at higher temperatures. Similar results can be expected from exoplanets.

%\red{CHECK: Hoeft J. Naturforsch A27, 703 (1972) is missing}

Laboratory spectroscopic studies of PN started with those of
\citet{33CuHeHe1.PN,33CuHeHe.PN}. Further experimental
investigations %into the spectrum
%of PN
were undertaken in the sub-millimetre, radiowave, microwave and
infra-red frequency ranges
\citep{71RaKlxx.PN,72HoTiTo.PN,72WyGoMa.PN,80CoPrxx.PN,81CoPrxx.PN,81MaLoxx.PN,81GhVeVa.PN,87VeGhIq.PN,96LeMeDu.PN,06CaClLi.PN}.
Some molecular and spectroscopic constants for PN computed \ai\ were
reported by
\citet{71RaKlxx.PN,72WyGoMa.PN,81GhVeVa.PN,87SaKrxx.PN,95AhHaxx.PN}.
%\red{Peterson and Dunning 2002 ??, Wang et al 2010 : CHECK!}.
Several attempts have been made to construct potential energy curves of PN
\citep{81MaLoxx.PN,83GrKaxx.PN,90WoRaxx.PN,92McLiCh.PN,93deFeKo.PN,03KeFrDi.PN}.
The vibrationally averaged dipole moments of $^{31}$P$^{14}$N for
$v=0, 1$, and $2$ were determined by \citet{71RaKlxx.PN} from
molecular beam electric resonance measurements.

The aim of this work is to produce accurate rotational-vibrational (ro-vibrational)
catalogues of transitions (so called line lists) for the two stable
isotopologues of PN, $^{31}$P$^{14}$N and $^{31}$P$^{15}$N, in their ground electronic states.
To this end an \ai\ dipole moment curve (DMC) is used in conjunction with an empirically
refined potential energy curve (PEC). Our line lists
%  comprise
% transition frequencies, Einstein coefficients, and various information on the
% lower and upper states \citep{jt528} and
should be complete enough to describe the absorption
of PN up to 5000~K.
This work is performed as part  the ExoMol project, which aims to provide
line lists for all the molecular transitions of importance in the atmospheres of
(exo-)planets and cool stars \citep{jt528}.
The ExoMol methodology has been already applied to a number
of diatomic molecules: BeH, MgH, CaH \citep{jt529},
SiO \citep{jt563}, NaCl and KCl \citep{jt583}, and AlO \citep{jtalo}.

% The line positions are accurate compared to the experimental data available in the literature.

\section{Solving the nuclear motion problem}
A spectroscopic line list comprises transition frequencies (i.e., line positions), Einstein coefficients
as well as energies, degeneracies and quantum numbers of the states involved in the transition
\citep{jt528}.  When combined with a partition function, a line list
can be used to produce spectra at a given temperature, both in absorption and emission.
Each record in the line list corresponds to a transition between two energy levels,
in this case ro-vibrational. These levels correspond to characteristic
nuclear motions of the molecule, and thus to compute  a line
list one has to solve the corresponding Schr\"{o}dinger equation and obtain
the energy levels required.

%\subsection{The Schr\"{o}dinger Equation}
The general form of the radial Schr\"{o}dinger equation for a diatomic molecule with
atoms $A$ and $B$ in a $^1\Sigma^+$ electronic state is given by
\begin{equation}
-\dfrac{\hbar^{2}}{2\mu}\dfrac{\mathrm{d}^{2}\psi_{v,J}(r)}{\mathrm{d}r^{2}}+\left[V(r) + \dfrac{\hbar^{2}}{2\mu r^2} J(J+1) \right]\psi_{v,J}(r) = E_{v,J} \psi_{v,J}(r), \label{eq6}
\end{equation}
where $V(r)$ is the potential energy curve,  $E_{v,J}$ and $\psi_{v,J}$ are,
respectively, the bound state eigenvalues and eigenvectors and $\mu$ is the reduced mass
given by
\begin{equation}
\mu = \frac{m_A m_B}{m_A+m_B}.
\end{equation}
In this work for all nuclear motion calculation we used the following (atomic) masses
\citep{12PfVeCz.db}: $m({^{31}}\mathrm{P}) = 30.97376200041$~u, $m({^{14}}\mathrm{N}) =
14.00307400461$~u and $m({^{15}}\mathrm{N}) = 15.00010889818$~u.

Einstein $\mathcal{A}$ coefficients in s$^{-1}$ for rovibrational transitions in $\Sigma$
electronic terms are given by \citep{05Bexxxx.method}
%LL The `times 10^{-36}' should be dropped if we do not specify that the dipole is in debye and Plank is in cgs units etc.
% NB that 1 D = 1e-18 statcoulomb*cm
\begin{equation}
\mathcal{A} = \frac{64 \pi^4 10^{-36}}{3 h } \frac{S(J\p, J\pp)}{2J\p+1} \, \tilde{\nu}^{3} \, |\langle\psi_{v\p,J\p}|M(r)|\psi_{v\pp,J\pp}\rangle |^{2}
\label{e:Aif}
\end{equation}
where $J\p$ and $J\pp$ are, respectively, the upper and lower rotational angular momentum
quantum numbers, $h=6.62606957 \times 10^{-27}$~erg s is Planck's constant in cgs units,
$S(J\p, J\pp) = \max(J\p, J\pp)$ is the H{\"o}nl-London rotational intensity factor,
$\tilde{\nu}$ is the transition wavenumber in \cm{} and $M(r)$ is the dipole moment curve
(DMC) in debyes.

Integrated absorbtion cross-sections $I$ (also referred to as integrated %(\tilde{\nu})
absorption coefficients of simply as ``line intensities'') in cm molecule$^{-1}$ can be obtained by
\begin{equation} %(\tilde{\nu})
I = \frac{1}{8 \pi c \, \tilde{\nu}^{2}} \, \frac{g_{\rm ns}(2 J\p +1)}{Q(T)} \ \mathcal{A} \
e^{-E^{\prime\prime}/ (k_B T)} \left( 1-e^{-h c \tilde{\nu}/ (k_B T) } \right)
\label{e:Ssigma}
\end{equation}
where $c = 2.99792458 \times 10^{10} $ cm s$^{-1}$ is the speed of light in cgs units,
${E}^{\prime\prime}$ is the lower state energy, $k_B$ is the Boltzmann constant, $T$ is
the temperature, $Q(T)$ is the partition functions, and $g_{\rm ns}$ is a nuclear spin
factor.

%$c_2 = 1.438 7770(13)$ cm~K is the second  radiation constant,

To solve the radial Schr\"{o}dinger equation \eqref{eq6} and compute
the energy levels, spectroscopic constants and Einstein coefficients
required we make use of the program LEVEL~8.0 \citep{lr07}. The PEC
and DMC are two prerequisites for these calculations and can be
obtained from first principles by solving the Schr\"{o}dinger equation
for the motion of the electrons. Since \ai\ PECs are not usually
capable of providing accuracies better than a few \cm, they are
commonly refined by fitting to experimental energies or transition
frequencies, subject to availability.  Conversely, the accuracy of
\ai\ DMC can be high enough to be competitive with, or even better
than, experiment \citep{jt156,jt509,jt573}. We use the DPotFit program
\citep{dpotfit} to empirically refine the \ai\ PEC by fitting to
experimental frequencies collected from different literature sources,
see Table~\ref{t:exp-data}.

\begin{center}
\begin{table}
\caption{Summary of experimental spectroscopic data available for PN used in our fits. }
\label{t:exp-data}
\begin{tabular}{lccccr} \hline
Ref. & Isotopes considered & max $v$ & max $J$ & Wavenumber range cm$^{-1}$ & Number of lines \\ \hline
%	\citet{06CaClLi.PN}    & $^{31}$P$^{14}$N & 4, $\Delta v = 1$ & 53 & 1217.2--1368.49 \red{CHECK without MICROWAVE} & 84 \\
	\citet{06CaClLi.PN}    & ${^{31}}$P$\,^{14}$N & 0 & 17 & 3.1--26.6 & 24 \\
	\citet{06CaClLi.PN}    & ${^{31}}$P$\,^{15}$N & 0 & 15 & 3.0--20.9  & 9 \\
	\citet{81MaLoxx.PN}    & ${^{31}}$P$\,^{14}$N & 4, $\Delta v = 1$ & 53 & 1217.22--1318.60 & 22  \\
	\citet{72WyGoMa.PN}    & ${^{31}}$P$\,^{14}$N & 4, $\Delta v = 0$& 8 & 3.1--12.5& 19  \\
	\citet{95AhHaxx.PN}    & ${^{31}}$P$\,^{14}$N & 1, $\Delta v = 1$& 33 & 1273.30--1368.49 & 62  \\
	\citet{81GhVeVa.PN}$^a$& ${^{31}}$P$\,^{14}$N & 11 & 0 & 1323.17 -- 13786.95$^b$ & 11 \\
 \hline
\end{tabular}
\mbox{}\\
$^a$ Vibrational band centres derived from the high resolution study of  $\leftexp{1}{\Pi}$--$\leftexp{1}{\Sigma^+}$ transition.\\
$^b$ Range for the lower ($\leftexp{1}{\Sigma^+}$) state term values \citep{81GhVeVa.PN}.
\end{table}
\end{center}

\section{Potential energy curve}
We produced in this work three potential curves for PN: an \emph{ab initio} one named
PEC-A; a semiempirical one named PEC-S which modifies the \emph{ab initio} PEC with a
single, physically-motivated, empirical parameter; and, finally, an empirical PEC-R
obtained by fitting the available experimental data. The three PECs are discussed in the
following sections.

% At this time affordable \ai\ methods do not allow reaching accuracies competitive with
% experiment (better than 0.01~\cm) for ro-vibrational line positions of molecules
% with more than about 5--10 electrons.
% The performance of several electronic structure methods and basis sets is illustrated
% in Table~\ref{t:en.levels}. All electronic structure calculations were performed with
% Molpro \cite{molpro2012}. The experimentally derived values were taken from \citet{95AhHaxx.PN} and
% to the zero-point energy from \citet{07Irxxxx.gen}. Vibrational ($J=0$) energy levels are
% known experimentally up to $v=11$ with a stated uncertainty of 0.001~\cm\ for
% $v=0$--4 and 0.05~\cm\ for $v=5$--11.
%
% With a view to improving the quality of the computed energy levels we considered
% semi-empirical and empirical potential energy curves which were adjusted to best fit
% known experimental data. These are discussed in the following sections.

\subsection{The \emph{ab initio} PEC-A}
\label{s:abinitio}

We computed PECs using the well-known CCSD (coupled cluster single and double
excitations) and CCSD(T) (coupled cluster single, double and perturbative triple
excitations) methods \citep{07BaMuxx.ai} as well as with CASSCF (complete active space
self consistent field), MRCI (internally contracted multi-reference configuration
interaction) and MRCI+Q (MRCI with renormalized Davidson correction) \citep{12SzMuGi.ai};
the +Q correction was computed using
the so-called `fixed' reference function, see \citet{12SzMuGi.ai} and the
MOLPRO manual for an explanation.  CASSCF and MRCI calculations used the full valence
reference space and all electrons apart from the phosphorus $1s$ were correlated.  We
used correlation-consistent basis sets of the Dunning family
\citep{89Duxxxx.ai,95WoDuxx.ai,02PeDuxx.ai} and in particular aug-cc-pwCV5Z and
aug-cc-pC6Z, for which we use the shorthand notation awc5z and ac6z.  We extrapolated the
energies using the formula $E_n = {E_\infty + A(n+\frac{1}{2})^{-4}}$, which is known to
perform very well \citep{12FePexx.ai}.  A relativistic correction curve was computed as
expectation value of the mass-velocity one-electron Darwin operator (MVD1) with the
MRCI/ac6z electronic wave function and was always included in the calculations. Energies
were computed from $r=2.10$~a$_0$ to $r=6.10$~a$_0$ (1.11 to 3.23~\AA) in steps of
0.05~a$_0$ and from $r=6.5$~a$_0$ to $r=16$~a$_0$ is steps of 0.5~a$_0$
using MOLPRO  \citep{12WeKnKn.methods}. Calculation of one MRCI
energy in the largest ac6z basis set took about 25~GB of disk and
3.5~hours on a single CPU.

\begin{table}
\begin{center}
\caption{Vibrational ($J=0$) energy term values obtained with different electronic structure methods and basis sets.
The second column, labelled `obs.', reports experimental (observed) values from
\citet{95AhHaxx.PN} and
\citet{07Irxxxx.gen}, other columns show observed minus calculated. The stated uncertainty of experimental levels
is 0.001~\cm\ for $v=0$--4 and 0.05~\cm\ for $v=5$--11.
The energy curve corresponding to the results in the last column is referred to as PEC-A in the text.
% CASSCF and MRCI calculations used the full-valence active space.
% All electrons apart from the phosphorus $1s$ were correlated.
In all cases a relativistic correction curve
computed with the MVD1 Hamiltonian at the MRCI/aug-cc-pV6Z level was included. See Section~\ref{s:abinitio} for detailed explanations.
All values are in \cm. \label{t:en.levels}}
\begin{tabular}{r r r r r r r r r}
\hline
\hline
\multicolumn{1}{l}{$v$} &  \multicolumn{1}{c}{obs.} & \multicolumn{7}{c}{obs. -- calc.} \\
\hline
                           &&   awc5z    &  awc5z     & awc5z    & awc5z   & awc5z    & ac6z      &  ac[56]z     \\  % &  awc5z     \\
                           &&  CCSD      &  CCSD(T)   &  CASSCF  &  MRCI   &  MRCI+Q  &  MRCI+Q   &  MRCI+Q      \\% &   +Q-rel   \\
 0 &      666.79      &    -40.28        &    -6.10   &   9.72   &   -5.85 &  -1.77   &   -2.55   &    -3.37     \\% &    -2.00   \\
 1 &     1989.94      &   -122.06        &   -18.78   &  28.93   &  -17.67 &  -5.37   &   -7.70   &   -10.16     \\% &    -6.06   \\
 2 &     3299.24      &   -205.74        &   -32.15   &  47.80   &  -29.71 &  -9.07   &  -12.97   &   -17.07     \\% &   -10.22   \\
 3 &     4594.64      &   -291.40        &   -46.24   &  66.34   &  -41.97 & -12.88   &  -18.35   &   -24.09     \\% &   -14.49   \\
 4 &     5876.14      &   -379.07        &   -61.11   &  84.57   &  -54.44 & -16.80   &  -23.84   &   -31.24     \\% &   -18.86   \\
 5 &     7143.83      &   -468.69        &   -76.61   & 102.66   &  -66.98 & -20.68   &  -29.31   &   -38.38     \\% &   -23.19   \\
 6 &     8397.42      &   -560.59        &   -93.07   & 120.35   &  -79.86 & -24.79   &  -35.02   &   -45.77     \\% &   -27.75   \\
 7 &     9637.02      &   -654.68        &  -110.35   & 137.83   &  -92.93 & -28.97   &  -40.82   &   -53.27     \\% &   -32.38   \\
 8 &   10~862.51      &   -751.11        &  -128.59   & 155.04   & -106.26 & -33.31   &  -46.79   &   -60.96     \\% &   -37.16   \\
 9 &   12~073.43      &   -850.38        &  -148.20   & 171.60   & -120.27 & -38.22   &  -53.36   &   -69.26     \\% &   -42.51   \\
10 &   13~270.68      &   -951.60        &  -168.27   & 188.49   & -134.00 & -42.74   &  -59.56   &   -77.22     \\% &   -47.47   \\
11 &   14~453.74      &  -1055.34        &  -189.33   & 205.30   & -147.92 & -47.35   &  -65.86   &   -85.31     \\% &   -52.51   \\
\hline
\hline
\end{tabular}
\end{center}
\end{table}

Some indicative results are reported in Table~\ref{t:en.levels}. From these it is clear
that errors in \emph{ab initio} energy levels are very large.  Let us first discuss the
results in the awc5z basis set. CCSD performs by far the poorest of the methods
considered; the (T) correction reduces CCSD errors by a factor about 6 but the
errors are still very large.  This seems to indicate that at the very least CCSDTQ
(coupled cluster with full triple and quadruple excitations) should be used for sub-\cm{}
accuracy, which is computationally unfeasible.  CASSCF energy levels are slightly worse
than CCSD(T) ones.  MRCI energy levels are slightly better than CCSD(T) ones, with errors
reduced by a factor 1.1 or so.  Finally, MRCI+Q improves considerably upon MRCI, reducing
errors by a factor about 3. They are still, however, considerably off, with absolute
errors in \cm{} progressively increasing with the vibrational quantum number as about
$4.15 \times (v+1/2)$.

Note that all methods, apart from CASSCF (which anyway is not expected
to produce quantitative accuracy),
produce energy levels which are too high,
suggesting the outer wall of the potential curve is rising too steeply. %relative PECs are too steep.
Increasing the basis set to ac6z leads to higher energies levels
and therefore to even larger errors;
basis set extrapolation inevitably leads to errors which are larger still.

We estimate the residual basis set incompleteness error in extrapolated ac[56]z energy
levels as half of the difference between ac[56]z and unextrapolated ac6z value; this
estimate is well approximated by the expression $0.84\times (v+1/2)$, indicating that residual
basis set errors are about five times smaller than the observed errors.
The relativistic MVD1 correction curve (included in all calculations reported in
Table~\ref{t:en.levels}) improves agreement with experiment; its effect on vibrational
energy levels $v=0$--11 can be expressed to within 0.04~\cm{} by $-1.38\times (v+1/2)$.
Further corrections not considered in this work such as higher relativistic effects due
to Gaunt, Breit or quantum electrodynamics corrections, as well as adiabatic and
non-adiabatic effects, are certainly considerably smaller than the included MVD1 curve
and therefore their inclusion at this stage would not lead to an improvement of the
observed residuals.
We reach therefore the conclusion that the observed large errors
in our best \emph{ab initio} results are mostly
due to incomplete treatment of electron correlation at the CCSD(T) or
MRCI+Q level.

% These results are in line with what observed with other
% molecules \red{LORENZO, REFERENCE?}: MRCI+Q in the full-valence active space performs
% considerably better than CCSD(T) but larger active spaces, currently computationally
% inaccessible, are needed for the spectroscopic accuracy.
Our highest-level \emph{ab initio} PEC is the MRCI+Q/ac[56]z/MVD1, based on extrapolation
of the awc5z and ac6z basis sets. This latter potential energy curve will be referred to
as PEC-A and leads to a  potential well depth $D_\mathrm{e} = 51~605$~\cm{}, which is
lower by 335~\cm{} than the (rather inaccurate) experimental value 51~940$\pm$170~\cm.

%% NOTE: The potential well depth of both A(r) and S(r) and obtained by extrapolation from r=6.1 a0 and
%% therefore may be rather off !!!
%%
%
It is worth noting that, almost certainly because of partial cancellation of
errors, best agreement with experiment is obtained \ai\ with the MRCI+Q/acw5z/MVD1 PEC,
at least for the range characterised by the available $v=0$--11 experimental data, which
are sensitive to the PEC in the range $r \approx 1.28$--1.80~\AA.

\subsection{The semi-empirical PEC-S}
We considered a simple one-parameter semi-empirical modification of PEC-A as a way of
improving its quality in a physically motivated way and with a view to provide a
reasonable behaviour in the high energy range, experimentally unexplored.  In particular,
we investigated simple scalings of the Davidson correction contribution to the energies.
The \emph{ab initio} MRCI+Q energies can be broken down into a sum of three
contributions: $E_\mathrm{MRCI+Q} = E_\mathrm{CASSCF} + E_\mathrm{corr} + E_\mathrm{+Q}$,
where $E_\mathrm{CASSCF}$ is the energy of the reference CASSCF wave function,
$E_\mathrm{corr}=  E_\mathrm{MRCI} - E_\mathrm{CASSCF}$ is the MRCI correlation energy
and $E_\mathrm{+Q}$ is the renormalized Davidson correction, which is given by
$E_\mathrm{+Q} = E_\mathrm{corr} (1-c_0^2)/c_0^2$ where $c_0$ is the expansion
coefficient of the CASSCF wave function in the MRCI one \citep{12SzMuGi.ai}. The Davidson
correction tries to approximate the effect of quadruple excitations neglected in the MRCI
calculation; in our calculation the $c_0$ expansion coefficient as a function of $r$ has
the value 0.966 for the near-equilibrium geometry $r=1.50$~\AA, then decreases linearly
up to about $r=2.0$~\AA, assumes a minimum value 0.963 at $r=2.30$~\AA{} and then slowly
grows up to a dissociation value of 0.965. The small variation of $c_0$ with geometry and
the fact $c_0 \approx 1$ show that the CASSCF wave function is indeed a good reference
function for all geometries. By contrast the expansion coefficient of the Hartree-Fock
reference in the MRCI expansions is 0.91 at $r=1.50$~\AA{} and then decreses
monotonically, is 0.5 at $r=2.56$~\AA{} and has an asymptotic value of about 0.28.
% We note that most of the error in the vibrational energies using MRCI+Q in the
% extrapolated [5-6] basis set  is probably due to the incomplete treatment of the electron
% correlation (see Table~\ref{t:en.levels}). Only values $v=0$ to $v=8$ were included in
% these scalings.
Eventually we chose %to use intead of the Davidson correction a semi-empirical
%correction of the form $E_\mathrm{corr} (1-c_0^2) + \alpha (1-c_0^2)^2$
%where $\alpha=7.29$ is a parameter adjusted to minimize residuals with experimental energies.
a simple direct scaling of the Davidson correction part of PEC-A by a factor
$\alpha=1.835$. The optimal value of $\alpha$ was found by fitting the experimental $J=0$
vibrational levels for $v=0$--11 reported in Table~\ref{t:en.levels}. Equivalently, PEC-S
can be written as V$_{\rm S}(r) = V_{\rm A}(r) + 0.835 \, E_\mathrm{+Q}(r)$, where
$V_{\rm A}(r)$ refers to PEC-A.

This one-parameter semi-empirical PEC, referred to as PEC-S below, reduces very
considerably the residuals with experiment of PEC-A and leads to agreements of better
than 0.30~\cm{} for to $v=0$--8, while last three levels $v=9$--11 show larger residuals
of the order of 1.5~\cm, see Table~\ref{t:obs-calc-vib}.
%  that $J=0$ levels for $v=0$ to $v=8$ can be fitted within
% 0.30~\cm{} using this form, while the last three levels $v=9$--11 show larger residuals.
The PEC-S leads to a potential well depth $D_\mathrm{e} = 51~742$~\cm{}, which is lower
by 198~\cm{} but compatible with the experimental value, 51~940$\pm$170~\cm. PEC-S was
used as the starting point for our subsequent empirical fit.

% Other, similar forms were also tested and gave similar results.
% Some indicative results of this kind of adjustment are reported in
% Table~\ref{t:obs-calc-vib}, taking as a basis the MRCI+Q-fix /AC[56]Z curve including the
% MVD1 relativistic correction and setting $\alpha = 7.29$. Another, simpler scheme which
% gives the same final results in terms of the `obs.-calc.' residuals  is to scale the
% +Q-fix correction by the factor 1.851.

% The semi-empirical correction can also be expressed approximately by a correction to the
% \emph{ab initio} curve having the form $V'(r)=-467 (r-r_e)$/\cm; this implies
% that at the outer limit relevant for the levels up to $v=11$, namely $r_\mathrm{max} =
% 3.7$~a$_0$, the correction due to electron correlation is of the order of 400~\cm.
%Plots of the corrections are reported in Appendix.

\subsection{The empirical PEC-R}\label{sec.PEC-R}
We considered a fully empirical potential energy function based on the Extended Morse
Oscillator (EMO) function \citep{lr07,84SuRaBo.method} as given by
\begin{equation}\label{e:EMO}
% V_{\rm EMO}(r)= D_{\rm e}[1-e^{-\beta_{\rm EMO}(r) (r-r_{\rm e})}]^2 ,
V(r)= D_{\rm e}[1-e^{-\beta(r) (r-r_{\rm e})}]^2, \ \quad \mathrm{where} \quad \beta(r) = \sum_{i=0}^{N_{\beta}}\beta_{i}\ \left( \frac{r^p - r_{\rm e}^p }{r^p + r_{\rm e}^p} \right)^i.
\end{equation}
% where
% \begin{equation} \label{eq14}
% %  \beta_{\rm EMO} (y^{r_{ref}}_{p}(r)) = \sum_{i=0}^{N_{\beta}}\beta_{i}y^{r_{ref}}_{q}(r)^{i} \label{eq14}
%  \beta(r) = \sum_{i=0}^{N_{\beta}}\beta_{i}\ \left( \frac{r^p - r_{\rm e}^p }{r^p + r_{\rm e}^p} \right)^i
% \end{equation}
The parameters to be fitted are the equilibrium
distance $r_{\rm e}$, the  potential well depth $D_{\rm e}$, the coefficients $\beta_i$ and, finally, the exponent $p$.
Note that one should constrain $p>0, \sum_i \beta_i > 0 $ in order to insure $V(+\infty)=D_\mathrm{e}$.
In our fits we truncated the expansion of $\beta(r)$ to three terms ($N_{\beta} = 2$) for $r < r_\mathrm{e}$. %; this introduces
% a discontinuity in the 5$^\mathrm{th}$ derivative
%  $y^{r_{ref}}_{p}(r)$ is the
% \v{S}urkus-variable \citep{84SuRaBo.method} given by
% \begin{equation}\label{e:surkus}
%  y^{r_{ref}}_{p}(r) = \dfrac{r^{p}-r_{ref}^{p}}{r^{p}+r^{p}_{ref}}
% \end{equation}
% with $p$ as a parameter.

We used DPotFit \citep{dpotfit} to refine the expansion parameters by fitting to the experimental
frequencies summarised in Table~\ref{t:exp-data}.  The main part of the experimental data
set includes pure rotational transitions recorded by \citet{06CaClLi.PN} and infrared transitions from
\citet{95AhHaxx.PN} which cover $v\le 4$. In order to improve the sampling for higher vibrational excitations,
we found it important to include the experimentally derived vibrational energies reported
by \citet{81GhVeVa.PN}, although with fitting weights smaller by a factor about 10 than
the rest of the data.  Since DPotFit is not able to fit to energies directly, we used
these energies to generate pseudo-transition frequencies, which allowed us to use data
from higher vibrational states, up to $v=11$, and obtain a much more reasonable fit at
the higher energy part of the PEC.

The values for the equilibrium distance $r_{\rm e}$ and dissociation energy $D_{\rm e}$
were fixed to the experimentally derived values of \citet{69Gixxxx.PN,06CaClLi.PN}. After
some experimentation the exponent $p$ in Eq.~(\ref{e:EMO}) was set to 4, which allowed us
to obtain an accurate fit to the higher energy levels.  Four expansion variables
$\beta_i$ were used in the final results.  Initial values for the parameters $\beta_i$
were set to those obtained for the scaled PEC-S. It should be noted, however, that the
final results are not very sensitive to the initial \ai\ curve \citep{jt563}.  The final
empirical parameters are given in Table~\ref{t:params}.

% One additional measure that can be taken is to include
% Born-Oppenheimer breakdown correction values especially for lighter molecule.
% In this case this proved unnecessary as we achieved an accurate fit without
% these terms.
We did not consider explicitly
non-adiabatic corrections %Born-Oppenheimer breakdown corrections
as these effects
are expected to have a small effect ($<0.05$~\cm)
on most transition frequences because of the relatively
large nuclear masses of PN and the absence of low-lying
excited electronic states.

% NOTE non-adiabatic can be estimated by the diff. (nuclear - atomic masses)
% Energy levels v=0-11 shift by up to 1.7 cm-1 from nuclear<->atomic.
% E(v+1)-E(v) shift by up to 0.17 cm-1.
% These are probably overestimates; if we assume core electron are attached
% to nuclei and instead od nuclear masses with compare atomic vs (nuclear+core electrons)
% one can halve those values. Even so, the differences in E(v+1)-E(v) are of up to ~0.08
%

%When including higher order expansion parameters $\beta_i$  in order to get
%the potential to behave at lower energies we needed to keep the radial equilibrium
%distance and the dissociation energy fixed to the values obtained for $N_{\beta}=0$
%\red{CONTRADICTS THE STATEMENT ABOVE, NEEDS CHECKING}.

Table \ref{t:obs-calc-vib} gives residuals for a selection of energy levels. These
results show that PEC-R generally improves on PEC-S in terms of agreement with experiment
and all levels are reproduced to better than 1~\cm.
% that whilst we do not have the accuracy of the lower excitations, all of these
% semi-empirical values are reproduced to better than 1~\cm.
Table~\ref{t:obs-calc} gives residuals for very accurate pure rotational transitions in
the microwave region for $^{31}$P$^{14}$N  and $^{31}$P$^{15}$N. The $^{31}$P$^{15}$N
microwave data in this table are the only available experimental data for this
isotopologue. Because of cancellation of errors calculated rotational transition
frequences are for rotational transitions are much more accurate than for rovibrational
ones. The \emph{ab initio} PEC-A gives the worse results, with a root-mean-square (rms)
error of 0.03~\cm; PEC-S improves very considerably over PEC-A and gives an rms error of
0.001~\cm. Finally, the empirical PEC-R gives the smallest rms of 7$\times 10^{-5}$~\cm.

Overall the empirical PEC-R reproduces the experimental transition wavenumbers used in
the fit (excluding the $v>4$ simulated transitions) with an rms error of 0.004~\cm.

\begin{table}
\begin{center}
\caption{Differences of experimental (`obs.') ro-vibrational energy levels with calculated ones (`calc.')
for three potential energy curves $X(r)$, $X=$A, S or R. PEC-A is an \emph{ab initio} curve based on MRCI+Q/ac[56]z/MVD1;
PEC-S is a one-parameter modification of PEC-A;
PEC-R is an 8-parameter empirical curve defined by Eq.~(\ref{e:EMO}) and table~\ref{t:params}.
Experimental energy levels are computed using the spectroscopic parameters by \citet{95AhHaxx.PN}
and include the zero point energy by \citet{07Irxxxx.gen}.
}
\label{t:obs-calc-vib}
\begin{tabular}{r r r r r r@{\hskip 0.4cm} r r r r r@{\hskip 0.4cm} r r r r}
\hline
\hline
\multicolumn{1}{l}{$v$}& \multicolumn{1}{c}{obs.}  &  \multicolumn{3}{c}{obs. -- calc.}&& \multicolumn{1}{c}{obs.}  &  \multicolumn{3}{c}{obs. -- calc.}&& \multicolumn{1}{c}{obs.}  &  \multicolumn{3}{c}{obs. -- calc.} \\
                         &&  \multicolumn{1}{c}{A}   &   \multicolumn{1}{c}{S}    &  \multicolumn{1}{c}{R}  &&&  \multicolumn{1}{c}{A}   &   \multicolumn{1}{c}{S}    &  \multicolumn{1}{c}{R}  &&&  \multicolumn{1}{c}{A}   &   \multicolumn{1}{c}{S}    &  \multicolumn{1}{c}{R}   \\
\hline
& \multicolumn{4}{c}{$J=0$} && \multicolumn{4}{c}{$J=10$} && \multicolumn{4}{c}{$J=20$} \\
       0  &   666.79  &      -3.37  &  0.02   &   0.00  &&     752.99     &   -3.53    &   0.03   &  0.00  &&   995.76    &  -3.99     &   0.04  &  0.00  \\
       1  &  1989.94  &     -10.16  &  0.07   &   0.00  &&    2075.53     &  -10.33    &   0.08   &  0.00  &&  2316.58    & -10.80     &   0.09  &  0.00  \\
       2  &  3299.24  &     -17.07  &  0.10   &   0.01  &&    3384.21     &  -17.24    &   0.11   &  0.00  &&  3623.54    & -17.72     &   0.12  &  0.00  \\
       3  &  4594.64  &     -24.09  &  0.10   &   0.00  &&    4679.00     &  -24.27    &   0.10   &  0.00  &&  4916.60    & -24.76     &   0.12  &  0.00  \\
       4  &  5876.14  &     -31.24  &  0.06   &  -0.01  &&    5959.88     &  -31.42    &   0.06   & -0.01  &&  6195.74    & -31.93     &   0.08  &  0.00  \\
       5  &  7143.83  &     -38.38  &  0.13   &   0.14  &&    7226.93     &  -38.58    &   0.12   &  0.12  &&  7461.00    & -39.16     &   0.07  &  0.07  \\
       6  &  8397.42  &     -45.77  &  0.04   &   0.18  &&    8479.94     &  -45.94    &   0.06   &  0.20  &&  8712.36    & -46.43     &   0.12  &  0.27  \\
       7  &  9637.02  &     -53.27  & -0.07   &   0.28  &&    9718.90     &  -53.47    &  -0.08   &  0.28  &&  9949.49    & -54.04     &  -0.09  &  0.28  \\
       8  & 10862.51  &     -60.96  & -0.29   &   0.38  &&   10943.77     &  -61.16    &  -0.28   &  0.39  && 11172.63    & -61.71     &  -0.27  &  0.43  \\
       9  & 12073.43  &     -69.26  & -1.02   &   0.08  &&   12154.06     &  -69.47    &  -1.02   &  0.09  && 12381.12    & -70.07     &  -1.04  &  0.12  \\
      10  & 13270.68  &     -77.22  & -1.33   &   0.36  &&   13350.08     &  -78.03    &  -1.93   & -0.22  && 13573.70    & -80.31     &  -3.61  & -1.84  \\
      11  & 14453.74  &     -85.31  & -1.68   &   0.77  &&   14533.22     &  -85.41    &  -1.56   &  0.92  && 14757.10    & -85.66     &  -1.21  &  1.35  \\
\hline
\hline
\end{tabular}
\end{center}
\end{table}

\begin{figure}
\epsfxsize=10.0cm \epsfbox{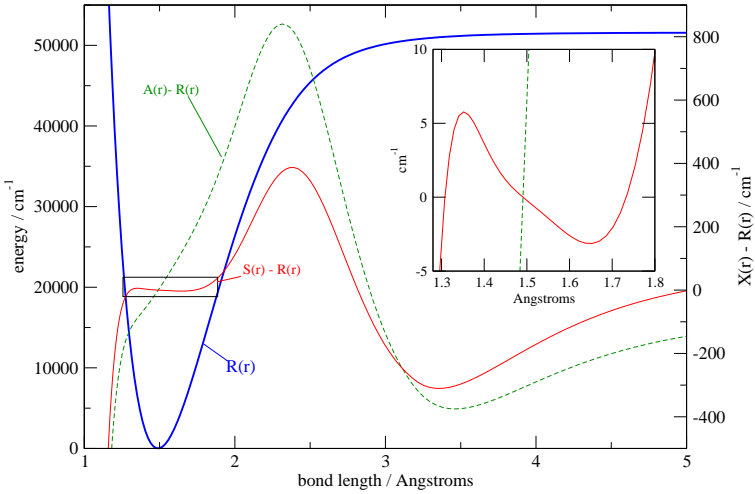}
\caption{The thick, blue curve is the 8-parameter empirical potential PEC-R defined by Eq.~(\ref{e:EMO}) and Table~\ref{t:params} (vertical scale given by the axis on the left hand side). The dashed, green curve is the difference between the \emph{ab initio} PEC-A
and PEC-R and the thin, red curve the difference between the one-parameter semi-empirical PEC-S
and PEC-R (vertical scale given by the axis on the right hand side). The inset on the right is a blow-up of the area $r = 1.28$--1.80~\AA{} contained in the black rectangle, corresponding to the range characterised by the available experimental data;
while within this range
PEC-S and PEC-R agree within about 10~\cm, outside this range differences can be as large as 400~\cm. \label{f:PEC}}
\end{figure}

Fig.~\ref{f:PEC} compares the refined PEC-R with the \ai\ PEC-A and the one-parameter
semiempirical PEC-S.  Due to the lack of experimental data for energies higher than about
14~500~\cm, only the range $r \approx 1.28$~\AA\ to $r \approx 1.80$~\AA\ of the
empirical PEC-R is well characterised.  As one can see from Fig.~\ref{f:PEC}, while
within this range PEC-R and PEC-S agree to about 10~\cm{} outside it the two PECs differ
by up to about 400~\cm. In particular, the peak visible for both PEC-A and PEC-S at $r
\approx 2.30$~\AA\ corresponds to the minimum of the $c_0$ curve (coefficient of the
CASSCF wave function in the MRCI expansion, see discussion in Section~\ref{s:abinitio}).
% In short, our \emph{ab
%   initio}-based curves $A(r)$ and $S(r)$ have a
% `bump' at $r \approx 2.30$~\AA\ due to the Davidson correction contribution.  %not true, it's not a `bump'
A more detailed study of the electronic structure of PN at long bond lengths would be
needed to ascertain whether this feature has a physical origin or is an artifact of our
calculation method, but this is beyond the scope of this work.  We can however note that
the difference in $J=0$ energy levels between PEC-S and PEC-R for $v > 11$ grow
approximately as $0.22 \times (v - 11)^2$~\cm\ up to $v = 47$ (PEC-R energy levels are the
higher ones) and reaches the maximum value of 210~\cm\ for that value of $v$.

Despite this large difference in energy levels transition frequences with $\Delta v = 1$
(the strongest) and $v > 11$ show maximum deviations between PEC-R and PEC-S of only up
to 10~\cm.

% It is reassuring that the shape
% of the refined PEC follows the \ai\ PEC up to the dissociation. This is despite the fact
% that the coverage of the experimental values is limited to 14~000~\cm.
% Furthermore, the \ai\ dissociation energy, 51~584~\cm, compares well with the experimentally derived
% value, 51~940~\cm.

\begin{center}	
\begin{table}
	\caption{Parameters defining the empirical potential surface PEC-R: expansion coefficients $\beta_i$,
    dissociation energy $D_{\rm e}$ in cm$^{-1}$, equilibrium bond length $r_{\rm e}$ in \AA, and parameter $p$, see eq.~(\ref{e:EMO}). For $r < r_\mathrm{e}$ the coefficients $\beta(3)$ and $\beta(4)$ are set to zero.}
    \label{t:params}
\begin{tabular}{lc}
  \hline\hline
  Parameter  &  Value  \\
  \hline
	     $D_{\rm e}/$\cm      &  51940 \\
		$r_{\rm e}/$ /AA      & 1.4908696082 \\
		 $\beta$(0)  & 2.2185248572 \\
		 $\beta$(1)  & 0.17749675548 \\
		 $\beta$(2)  & 0.16962236112 \\
		 $\beta$(3)  & 0.14242140368 \\
		 $\beta$(4)  & 0.22164327612 \\
		 p         & 4                 \\
\hline
	\end{tabular}
\end{table}
\end{center}

\begin{table}
\caption{Differences of experimental \protect\citep{72WyGoMa.PN,06CaClLi.PN}.
(`obs.') pure rotational transitions with calculated ones (`calc.')
for three potential energy curves PEC-$X$, $X=$A, S or R. All transitions have $\Delta v =0$ and $J'' = J'+1$. Energies are in \cm. \label{t:obs-calc}}
\begin{tabular}{r r r r r r r}
\hline
$v$ & $J\p$& obs.            & \multicolumn{3}{c}{(obs. - calc.) $\times 10^{5}$ } \\
    &      &                 & $A$ & $S$ & $R$\\
\hline
\multicolumn{6}{c}{P$^{14}$N} \\                                           %     PEC-R
  0 &   0  &       1.567427  &   -295.5    &    10.9   &  -0.5 \\         %    1.567437
  0 &   1  &       3.134828  &   -590.9    &    21.8   &  -1.0 \\         %    3.134832
  0 &   2  &       4.702176  &   -886.5    &    32.6   &  -1.6 \\         %    4.702192
  0 &   3  &       6.269446  &   -1182.0   &   43.5    & -2.2  \\         %    6.269468
  0 &   4  &       7.836611  &   -1477.6   &   54.3    & -2.7  \\         %    7.836638
  0 &   5  &       9.403643  &   -1773.5   &   65.0    & -3.5  \\         %    9.403678
  0 &   6  &      10.970523  &   -2069.0   &   76.1    & -3.8  \\         %   10.970561
  0 &   7  &      12.537217  &   -2364.8   &   86.9    & -4.4  \\         %   12.537261
  0 &  10  &      17.235939  &   -3252.9   &   119.4    &-6.0   \\         %   17.236000
  0 &  11  &      18.801641  &   -3549.0   &   130.4    &-6.4   \\         %   18.801705
  0 &  12  &      20.367025  &   -3845.6   &   141.1    &-7.2   \\         %   20.367097
  0 &  13  &      21.932071  &   -4142.2   &   151.9    &-7.7   \\         %   21.932148
  0 &  14  &      23.496750  &   -4439.0   &   162.7    &-8.3   \\         %   23.496833
  0 &  15  &      25.061036  &   -4736.0   &   173.3    &-9.0   \\         %   25.061127
  0 &  16  &      26.624905  &   -5033.1   &   184.2    &-9.5   \\         %   26.625000
\mbox{}\\
  1 &   1  &       3.112628  &   -606.1    &    21.6   &  -4.7  \\         %    3.112669
  1 &   2  &       4.668876  &   -909.3    &    32.3   &  -7.1  \\         %    4.668947
  1 &   3  &       6.225046  &   -1212.3   &   43.2    & -9.4   \\         %    6.225140
  1 &   4  &       7.781110  &   -1515.6   &   53.9    & -11.8   \\         %    7.781228
  1 &   5  &       9.337046  &   -1818.6   &   64.9    & -13.9   \\         %    9.337186
  1 &   6  &      10.892821  &   -2122.0   &   75.6    & -16.4   \\         %   10.892984
\mbox{}\\
  2 &   2  &       4.635485  &   -932.7   &    31.7   &  -7.9  \\         %    4.635564
  2 &   3  &       6.180523  &   -1243.8   &   42.2    & -10.6   \\         %    6.180629
  2 &   4  &       7.725463  &   -1554.2   &   53.4    & -12.5   \\         %    7.725588
\mbox{}\\
  3 &   3  &       6.135871  &   -1276.5   &   40.3    & -6.1   \\         %    6.135933
  3 &   4  &       7.669644  &   -1595.4   &   50.7    & -7.4   \\         %    7.669717
\mbox{}\\
  4 &   3  &       6.091073  &   -1310.7   &   37.2    & 2.6   \\         %    6.091047
\hline
\multicolumn{6}{c}{P$^{15}$N}\\
  0 &     1  &     2.991601  &  -561.6      &   22.9      &   1.1  \\         %    2.991600
  0 &     2  &     4.487339  &  -842.8      &   34.0      &   1.4  \\         %    4.487300
  0 &     7  &    11.964490  &  -2248.2     &  90.6      &  3.6   \\         %   11.964500
  0 &     8  &    13.459445  &  -2529.4     &  102.0     & 4.2    \\         %   13.459400
  0 &    10  &    16.448686  &  -3092.3     &  124.7     & 5.2    \\         %   16.448600
  0 &    11  &    17.942926  &  -3373.8     &  136.1     & 5.7    \\         %   17.942900
  0 &    12  &    19.436878  &  -3655.7     &  147.2     & 6.0    \\         %   19.436800
  0 &    13  &    20.930522  &  -3937.6     &  158.6     & 6.5    \\         %   20.930500
  0 &    14  &    22.423832  &  -4219.6     &  169.8     & 7.0    \\         %   22.423700
\hline
\end{tabular}
\end{table}

\subsection{Dipole moment curve (DMC)}
\begin{figure}
\centering
\epsfxsize=10.0cm \epsfbox{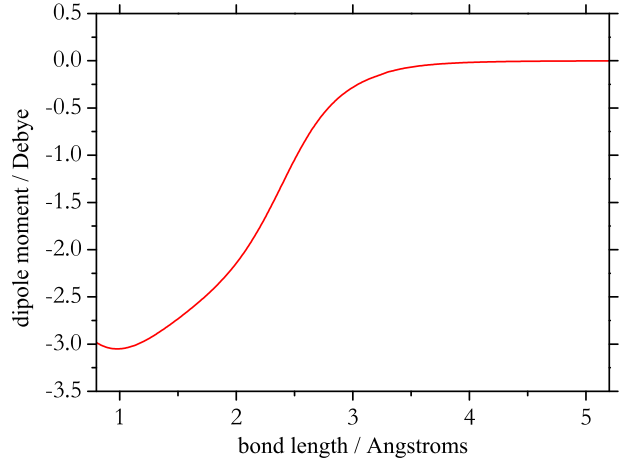}
\caption{\label{f:DMC}  The \ai\ MRCI+Q/aug-cc-pCV6Z dipole moment curve (DMC) of PN.}\end{figure}

There appears to be no full DMC for the ground electronic state of PN available in the
literature.  Furthermore, there are no experimental absolute transition intensities, so
it is difficult to judge the accuracy of a DMC.
% Therefore the only procedure is to choose
% the \ai\ method used for the dipole moments based on the performance of the corresponding
% PEC in terms of the ro-vibrational energies. As discussed in Section~\ref{s:abinitio} and
% shown by Table~\ref{t:en.levels}, among all \emph{ab initio} methods considered the
% MRCI/aug-cc-pwC5Z+Q/MVD1 level of theory gives best agreement with experiment.
We therefore used the highest level of \emph{ab initio} theory considered in this work,
MRCI+Q / aug-cc-pCV6Z, to compute a new DMC.
% We therefore used this method and computed
Dipoles were computed as the derivative of the MRCI+Q energy
with respect to an external electric field along the internuclear axis
for vanishing field strength \citep{jt475}; we used field
strengths $\pm 5\times 10^{-4}$~a.u. and
computed dipoles on a grid $r = 2.1$ to $6.1$~a$_0$ in steps
of 0.05~a$_0$. The corresponding DMC is shown in Fig.~\ref{f:DMC}.
This DMC has an equilibrium value of 2.739~D, which is close to the
value of the dipole moment, 2.7465$ \pm 0.0006$~D,
determined experimentally by \citet{71RaKlxx.PN}.

\subsection{Computing the line list}
The empirical PEC and the \ai\ DMC described above were used to generate the line lists
for $^{31}$P$^{14}$N and $^{31}$P$^{15}$N with LEVEL~8.0.  The integration range was
chosen as $r = [1.0, 10.0]$~\AA\ and the grid comprised 45~000 points.
We estimated the numerical error in computed energy levels to be less than 0.0005~\cm.
Table \ref{t:linelist} summarises the results, showing the maximal values of $v$, $J$ as well
as the wavenumber range considered in this work, which is far more extensive than any
previous study. However, as discussed in Section~\ref{sec.PEC-R}, our empirical potential
curve is not well characterised for energies higher than about 14~500~\cm{} and therefore
transition frequences involving energy levels with energies higher than
$\approx$~14~500~\cm{} may have large errors of up to a few hundreds of \cm.

Following \citet{jt548} our line lists consist of an `Energy' and a `Transition'
file, extracts from which are given in Table~\ref{t:Energy-and-transition-file}.
The full line list can be downloaded from the Exomol web site  \url{www.exomol.com}
or from the VizieR service from the web site of the Strasbourg Astronomical Data
Center \url{cds.u-strasbg.fr}. A small computer program which uses these files to compute
spectra is given in the supplementary material.

\begin{table}
\caption{Summary of our line lists.}
\begin{tabular}{lcr}
\hline\hline
         & $^{31}$P$^{14}$N & $^{31}$P$^{15}$N \\ \hline
         Max $v$ & 66 & 68  \\
         Max $J$ & 357 & 366 \\
         Max $\nu$ cm$^{-1}$ & 51928.7  & 51937.0 \\
         number of lines & 692019 & 743114 \\
         number of energies & 13949 & 14623 \\
         \hline
\end{tabular}
\label{t:linelist}
\end{table}

\begin{table}
\caption{Sample extracts from the energy and transition files for $^{31}$P$^{14}$N. The energy files contain 13~949 entries for $^{31}$P$^{14}$N and 14~623  for $^{31}$P$^{15}$N, while the transition files contain  692~019  entries for $^{31}$P$^{14}$N and 743~114   for $^{31}$P$^{15}$N.}
\label{t:Energy-and-transition-file}
\begin{tabular}{rrrrr r@{\hskip 0.8cm} r r r }
\cline{1-5}\cline{7-9}
\multicolumn{5}{c}{\texttt{Energy file}} && \multicolumn{3}{c}{\texttt{Transition file}}\\
     $N$ & \multicolumn{1}{c}{$\tilde{E}$} &  $g$    & $J$      & $v$    && $I$ & $F$ & \multicolumn{1}{r}{$A_{\rm if}$} \\
\cline{1-5}\cline{7-9}
            1  &   0.000000  &    4    &     0    &     0  &&         2   &           1   &        3.0148E-06  \\
            2  &   1.567437  &   12    &     1    &     0  &&         3   &           2   &        2.8941E-05  \\
            3  &   4.702269  &   20    &     2    &     0  &&         4   &           3   &        1.0465E-04  \\
            4  &   9.404461  &   28    &     3    &     0  &&         5   &           4   &        2.5722E-04  \\
            5  &  15.673929  &   36    &     4    &     0  &&         6   &           5   &        5.1376E-04  \\
            6  &  23.510567  &   44    &     5    &     0  &&         7   &           6   &        9.0135E-04  \\
            7  &  32.914245  &   52    &     6    &     0  &&         8   &           7   &        1.4470E-03  \\
            8  &  43.884806  &   60    &     7    &     0  &&         9   &           8   &        2.1778E-03  \\
            9  &  56.422067  &   68    &     8    &     0  &&        10   &           9   &        3.1207E-03  \\
           10  &  70.525819  &   76    &     9    &     0  &&        11   &          10   &        4.3027E-03  \\
           11  &  86.195825  &   84    &    10    &     0  &&        12   &          11   &        5.7506E-03  \\
           12  & 103.431825  &   92    &    11    &     0  &&        13   &          12   &        7.4913E-03  \\
\cline{1-5}\cline{7-9}
\mbox{}\\
\multicolumn{5}{l}{$N$:   State counting number.}      &&   \multicolumn{3}{l}{$I$: Upper state counting number.} \\
\multicolumn{5}{l}{$\tilde{E}$: State energy in \cm.}  &&   \multicolumn{3}{l}{$F$:      Lower state counting number.}\\
\multicolumn{5}{l}{$g$: State degeneracy.}             &&   \multicolumn{3}{l}{$A_{\rm if}$:  Einstein $\mathcal{A}$ coefficient in s$^{-1}$.}\\
\multicolumn{5}{l}{$J$:   State rotational quantum number.} \\
\multicolumn{5}{l}{$v$:   State vibrational quantum number.} \\
\end{tabular}
\end{table}
% $N$:   State counting number.
% $\tilde{E}$: State energy in \cm.
% $g$: State degeneracy.
% $J$:   State rotational quantum number.
% $v$:   State vibrational quantum number.
% $I$: Upper state counting number.
% $F$:      Lower state counting number.
% $A_{\rm if}$:  Einstein $\mathcal{A}$ coefficient in s$^{-1}$.

\subsection{Partition Function}
The partition functions, $Q(T)$, for $^{31}$P$^{14}$N and $^{31}$P$^{15}$N were
calculated by direct summation over the energy levels using
\begin{equation}
	Q(T)=g_{\rm ns}\sum_{i=0}^{n} \left( 2 J_i+1 \right ) e^{-E_{i} / (k_B T)} ,
\label{e:pf}
\end{equation}
where $J_i$ is the rotational angular momentum quantum number and $E_{i}$ is the energy
of the of state $i$. Partition functions can be used to compute a variety of
thermodynamic data.

ExoMol follows the
HITRAN convention of explicitly including the full atomic nuclear spin
in the molecular partition function \citep{03FiGaGo}. In our case
$g_{\rm ns} = 4$ for $^{31}$P$^{14}$N and $g_{\rm ns} = 6$ for $^{31}$P$^{15}$N.

The  $^{31}$P$^{14}$N partition function is illustrated in Table~\ref{t:pf}, where it is
compared to the values obtained from the parameters provided by
\citet{81Irxxxx.partfunc}. The agreement is good especially considering a rather limited
level of accuracy of the model used by \citet{81Irxxxx.partfunc}. At lower temperatures
our values do deviate somewhat from those derived by \citet{81Irxxxx.partfunc}, which is
expected  since the latter is only designed for $T>1000$~K. In total we have considered
temperatures from 10~K--5000~K in this work, and a full table of values for the partition
function values on a fine grid for the two isotopologues can be found in the
supplementary material. We also computed partition functions using energy levels obtained
from the semi-empirical PEC-S to estimate the error in our computed $Q(T)$. The
conclusion is that our partition functions should have a relative error of less than 0.01\%,
i.e. they are correct to four significant digits.
%Figure~\ref{f:Cp} displays values of the specific heat of $^{31}$P$^{14}$N  in units of the gas constant $R$ computed using our partition function.

%\begin{figure} [H]
%\caption{ The specific heat, $C_p$ in the units of the Gas constant $R$, of $^{31}$P$^{14}$N as function of temperature.}
%\label{f:Cp}
%\epsfxsize=10.0cm \epsfbox{Cp.eps}
%\end{figure}

\begin{table}
\caption{Tabulations of the partition function for $^{31}$P$^{14}$N at given temperatures in kelvin compared to the values by \citet{81Irxxxx.partfunc}.
The latter  were scaled by the $g_{\rm ns}=4$ factor. The full table can be found in the supplementary material. }
\label{t:pf}
\begin{tabular}{rrr}
\hline
$T$ & This work & \citet{81Irxxxx.partfunc}\\
\hline
     100   &       356.2   &       515.6   \\
     200   &       711.2   &       848.3   \\
     300   &      1068.2   &      1156.7   \\
     400   &      1434.1   &      1487.1   \\
     500   &      1817.9   &      1848.6   \\
     600   &      2227.0   &      2244.5   \\
     700   &      2666.1   &      2676.0   \\
     800   &      3138.3   &      3143.9   \\
     900   &      3645.3   &      3648.7   \\
    1000   &      4188.4   &      4190.8   \\
    1500   &      7468.0   &      7470.7   \\
    2000   &     11717.8   &     11719.8   \\
    2500   &     16964.4   &     16961.2   \\
    3000   &     23229.7   &     23215.1   \\
    3500   &     30535.8   &     30500.1   \\
    4000   &     38905.7   &     38834.1   \\
    4500   &     48363.9   &     48234.7   \\
    5000   &     58936.7   &     58719.8   \\
\hline
\end{tabular}
\end{table}

\subsection{Examples of ro-vibrational spectra}
%The intensity of a line is a measure of its strength at a given temperature. In emission
%the line intensity (emissivity) is given by (\red{UNITS}) \red{CHECK}
%\begin{equation}
%I(\mu_{ij})=\dfrac{hcA_{ij}E_{ij}g_{i}g_{ns}}{4{\pi}Q(T)}(e^{\frac{-E_{i}}{kT}}).
%\end{equation} \label{eq9}
%where ${\Delta}E_{ij}$ is the difference in energy between levels and $A_{ij}$ is the
%Einstein coefficient, $Q(T)$ is the partition function, and $g_{\rm ns}$ is the nuclear spin
%factor.
\begin{figure}
\label{f:absorp}
\epsfxsize=10.0cm \epsfbox{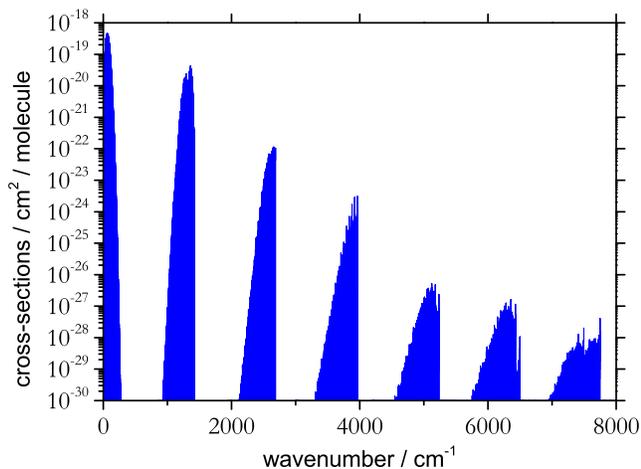}
\caption{Overview of the absorption spectrum of $^{31}$P$^{14}$N at a temperature of 1200 K.}
\end{figure}

\begin{figure}
\label{f:exp}
\epsfxsize=10.0cm \epsfbox{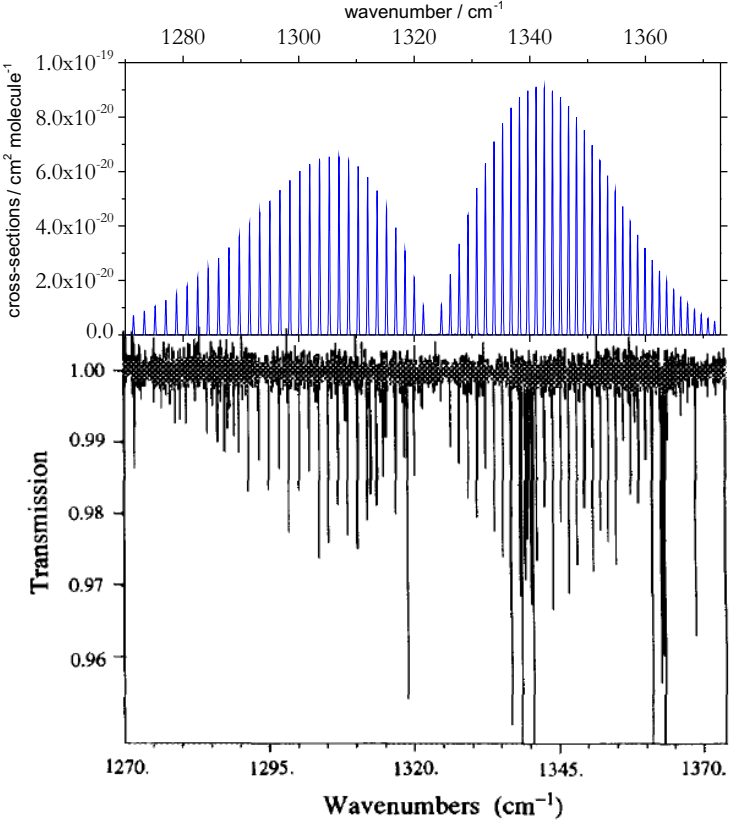}
\caption{Comparison between the experimental spectrum of $^{31}$P$^{14}$N by \protect\citet{95AhHaxx.PN}
and this work, at room temperature.  Our work is given in the upper panel
and has intensities in absolute units. The experimental data is given as a transmission
spectrum and shows many strong features due to water. (Reprinted from Ref.~\protect\citep{95AhHaxx.PN}. Copyright
1995, with permission from Elsevier.)}
\end{figure}

\begin{figure}
\epsfxsize=10.0cm \epsfbox{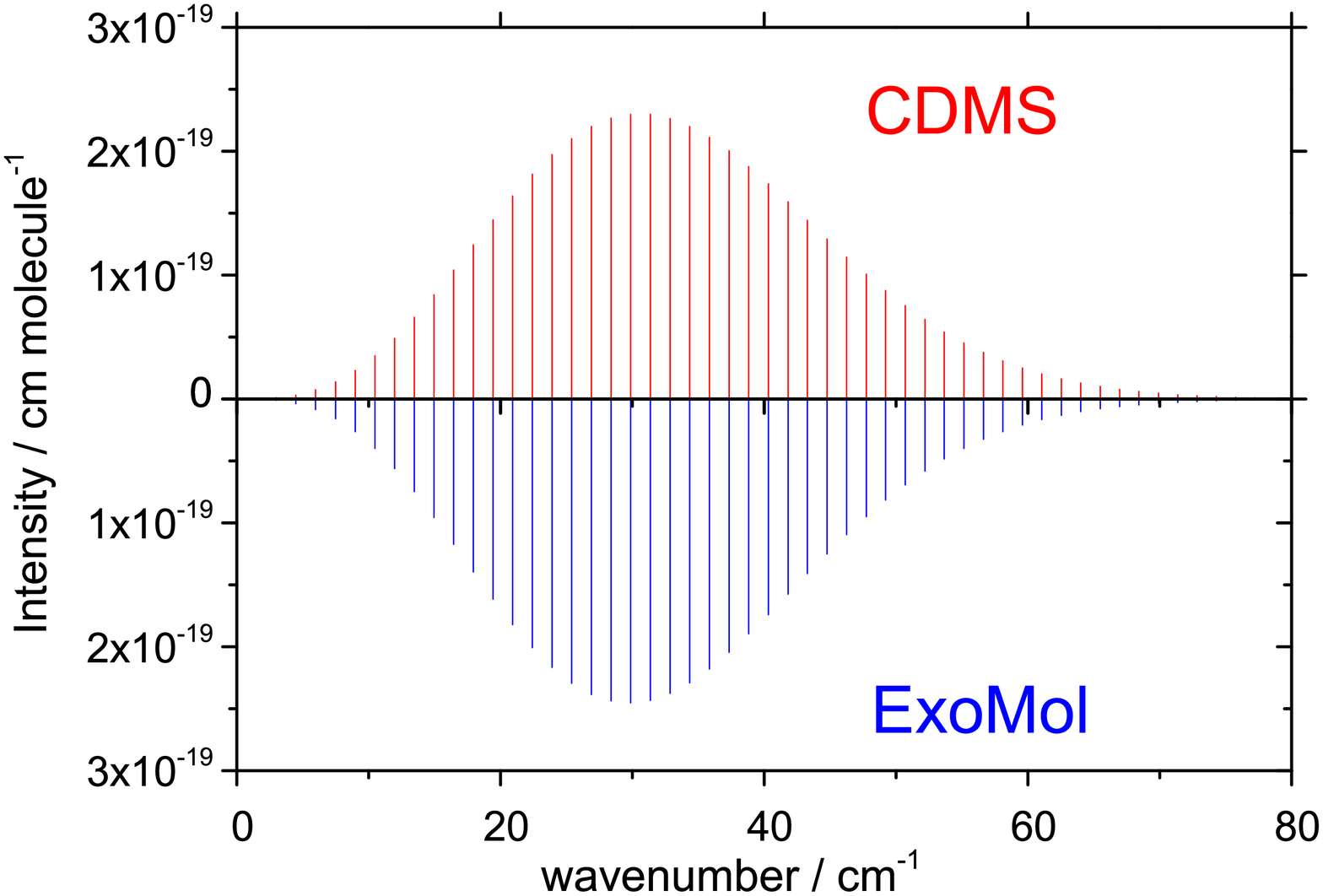}
\caption{Comparison between the room-temperature $^{31}$P$^{14}$N microwave spectrum from the
Cologne Database for Molecular Spectroscopy (CDMS) \citep{cdms} and this work.}
\label{f:CDMS}
\end{figure}

We used the new line list of $^{31}$P$^{14}$N to compute a number of absorption spectra
in the form of integrated cross-sections as described by \citet{jt542}. Figure \ref{f:absorp}
shows the absorption spectrum of $^{31}$P$^{14}$N at $T=1~200$~K.
% as cross-sections  where the Doppler profile was used.
Only $^{31}$P$^{14}$N is presented as the difference with spectra for $^{31}$P$^{15}$N
can only be distinguished at high resolution.
% so looking at
%Fig.~\ref{f:xxxx} for example, what appears to be a linear decay, is actually an
%exponential decay of intensity. If we look at the same temperature but in absorption
%(Fig.~\ref{f:xzzz}), we see that there is absorption at much higher wavenumbers, and it
%does not show the same level of decay as the equivalent emission diagram as affected by
%the Boltzmann population distribution. \red{CHECK}
Figure \ref{f:exp} compares a spectrum of $^{31}$P$^{14}$N recorded by
\citet{95AhHaxx.PN} with the spectrum from this work. Due to pollution of the
experimental spectrum with water lines, the spectra are far from identical; however they
show similar structure, with the $P$ and $R$ branches clearly visible in both spectra.
This demonstrates the accuracy of our results. Another illustration is presented in
Fig.~\ref{f:CDMS}, where the microwave spectrum of $^{31}$P$^{14}$N collected in the
Cologne Database for Molecular Spectroscopy (CDMS) \citep{cdms} compared to our spectrum.
The agrement is excellent.

%\begin{figure}[H]
	%\includegraphics[scale=0.9]{thesis9}
%	\caption{The complete line list for $^{31}$P$^{14}$N at 300K in emission}
%	\end{figure}
	
%	\begin{figure}[H]
%		%\includegraphics[scale=0.9]{thesis8}
%		\caption{P$^{14}$N at 300k in absorption}
%		\end{figure}
	
%	\begin{figure}[H]
%		%\includegraphics[scale=0.9]{thesis6}
%		\caption{The complete line list for $^{31}$P$^{14}$N
%		 at 1000K in emission}
%		\end{figure}
		
%	\begin{figure}[H]
%		%\includegraphics[scale=0.9]{thesis7}
%		\caption{The complete line list for $^{31}$P$^{14}$N at 1000K in absorption}
%		\end{figure}

\section{Discussion and Conclusion}
We produced three new potential energy curve (PEC) for PN: an \emph{ab initio} PEC-A, a
one-parameter semi-empirical PEC-S and a fully empirical 8-parameter PEC-R. The produced
PECs are applicable to both the $^{31}$P$^{14}$N and $^{31}$P$^{15}$N isotopologues. The
empirical PEC-R reproduces all known $^{31}$P$^{14}$N
experimental data up to $v=4$  with a typical error
of less than 0.01~\cm\ and is expected to be the most accurate of the three.
Transition frequencies for the minor isotopologue $^{31}$P$^{15}$N are expected to be somewhat less accurate.
Experimental
pure rotational transitions for both isotopologues with $v \leq 4$ \citep{06CaClLi.PN,95AhHaxx.PN}
in the microwave region are reproduced within $10^{-4}$~\cm.
Levels with $v=5$ to  $v=11$ are expected to have
larger errors, in the range 0.1 to 1.0~\cm.
The accuracy of semiempirical PEC-S should be
in most cases not much worse than the quoted values for PEC-R. Errors in the \emph{ab
initio} PEC-A are larger by a factor 10--100. Levels with $v > 11$ are expected to have
very large errors for all PECs, in the range 10--200~\cm; however, transition frequences
with $v > 11$ and $\Delta v = 1$ should be accurate to $\approx $5~\cm. Further experimental and
theoretical studies are needed to improve the accuracy for high $v$'s.

An \emph{ab initio} dipole moment was produced and, together with the PEC-R, used to
compute line lists (comprising transition line positions and Einstein coefficients) for
$^{31}$P$^{14}$N and $^{31}$P$^{15}$N. Details of the line lists are reported in
Table~\ref{t:linelist}. The line list cover all rotational-vibrational levels up to
dissociation but, as discussed above, transitions involving $ v > 11$ may be seriously in
error.

% Einstein coefficients were also calculated for all transitions.  There are
% no experimentally derived values to compare to, so again we have to assume that our
% level of accuracy has been maintained throughout calculations.  The Einstein coefficients
% were calculated using an \ai\ DMC calculated in this work as well.  This should be
% reasonably accurate and consequently lead to accurate values for the Einstein
% coefficients for both isotopologues.  Once again however, this would benefit from
% experimental verification.

Partition functions accurate to about 0.01\%
were also calculated for a range of temperatures for both isotopologues.
Work done previously by \citet{81Irxxxx.partfunc} shows a great deal of agreement with
our partition function, with the exception of temperatures below 1000~K where Irwin's
results are not valid.

\section{Supplementary material}
The Supplementary material to this paper includes the line lists and partition functions
of $^{31}$P$^{14}$N and $^{31}$P$^{15}$N, as well as the  PECs and DMCs used in these
calculations together with the LEVEL~8.0 and DPotFit input/output files, and a Fortran program which uses the
state and transition files to compute
spectra. The line list
are available from the Exomol web site  \url{www.exomol.com} or from the VizieR service
from the web site of the Strasbourg Astronomical Data Center via
\url{http://cdsarc.u-strasbg.fr/cgi-bin/VizieR?-source=J/MNRAS/xxx/yy}.

%\url{http://cdsarc.u-strasbg.fr/cgi-bin/VizieR?-source=J/MNRAS/xxx/yy}
%and via \url{www.exomol.com}.

%To conclude, i have produced two ro-vibrational line lists, for the ground states of $^{31}$P$^{14}$N and $^{31}$P$^{15}$N
%respectively.  The line list obtain a good level of accuracy and are the most comprehensive work done on these
%molecules to date, and should prove useful in further observational work.

\section*{Acknowledgements}

This work is supported by  ERC Advanced Investigator Project 267219.

\bibliographystyle{mn2e}
%\bibliography{journals_astro,PN,jtj,sy,methods,additional,exogen,phdthesis,h2s}

\end{document}